\documentclass{elsart}

\usepackage{graphicx}

\usepackage{color}

\usepackage{amssymb}

\newcommand{\norm}{\frac{2}{(2\pi)^3}}
\newcommand{\qd}[1]{\left[ #1 \right]}
\newcommand{\td}[1]{\left( #1 \right)}

\newcommand{\itg}[1]{\norm \int d^3k f_{#1}(k)}
\newcommand{\ienne}[1]{\itg{n} #1}
\newcommand{\ipi}[1]{\itg{p} #1}

\newcommand{\itau}{\itg{\tau}g(k,\Lambda)}

\newcommand{\rz}{\rho_{_0}}
\newcommand{\ra}{\td{ \frac{\rho}{\rho_{_0}} }}

\newcommand{\umd}[1]{ \td{ \frac{1}{2}+x_{#1} } }

\newcommand{\inew}[1]{\mathcal{I}_#1 }

\begin{document}

\begin{frontmatter}

\title{Transport properties of isospin effective mass splitting}
\author[catania]{J. Rizzo},
\author[catania]{M. Colonna},
\author[catania]{M. Di Toro\thanksref{dit}}
 and
\author[texas]{V. Greco}

\thanks[dit]{ditoro@lns.infn.it}

\address[catania]{Laboratori Nazionali del Sud INFN, I-95123 Catania, Italy\\
 Physics \& Astronomy Dept., Univ. of Catania }
\address[texas]{Cyclotron Institute, Texas A\& M University, 
College Station, USA}  
\begin{abstract}
We investigate in detail the momentum dependence ($MD$) of the effective 
in medium Nucleon-Nucleon ($NN$) interaction in the isovector channel.
We focus the discussion on transport properties of the expected
neutron-proton ($n/p$) effective mass splitting at high isospin density.
We look at observable effects from collective flows in 
Heavy Ion Collisions ($HIC$) of charge asymmetric nuclei at intermediate
energies. Using microscopic kinetic equation simulations nucleon
transverse and elliptic collective flows in $Au+Au$ collisions
are evaluated. In spite of the reduced charge asymmetry of the interacting
system interesting $isospin-MD$ effects are revealed. Good observables,
particularly sensitive to the $n/p$-mass splitting, appear to be
the differences between neutron and proton flows. The importance of
more exclusive measurements, with a selection of different bins of
the transverse momenta ($p_t$) of the emitted particles, is stressed.
In more inclusive data a compensation can be expected from different
$p_t$-contributions, due to the microscopic $isospin-MD$ structure of the
nuclear mean field in asymmetric matter.
\end{abstract}
\end{frontmatter}
\section{Introduction}
A key question in the physics of
unstable nuclei is the knowledge of the $EOS$ for asymmetric nuclear
matter away from normal conditions. 
We remind again the effect of the symmetry term at low densities
on the neutron skin structure, while the knowledge in
high densities region is crucial for supernovae dynamics
and neutron star cooling. The paradox is that while we
are planning second and third generation facilities for
radioactive beams our basic knowledge of the symmetry
term of the nuclear Equation of State $EOS$ is still extremely poor.
Effective interactions are obviously tuned to symmetry properties
around normal conditions and any extrapolation can be quite
dangerous. Microscopic approaches based on realistic $NN$
interactions, Brueckner or variational schemes, or on effective
field theories show a quite large variety of predictions,
see the discussions in \cite{iso01,dit03} and refs. therein.

In this paper we look at the mean field momentum dependence ($MD$) in the
isospin channel and study its dynamical effects 
in heavy ion collisions at intermediate energies.
This represents a completely open problem as we can quickly realize
from a simple analysis of the effective Skyrme-like non-relativistic
forces. Let's look at the local and non-local contributions 
to the potential symmetry term of various Skyrme interactions, $SIII$,
 $SGII$, $SKM*$ (see \cite{kri80} and refs. therein) and the more recent
Skyrme-Lyon forms, $SLy230a$ and $Sly230b$ (or $Sly4$), \cite{lyon97}.
We can clearly see a sharp transition
from the previous Skyrme forces to the Lyon parametrizations, with an
almost inversion of the signs of the two contributions. The important
repulsive non-local part of the Lyon forces leads to a completely
different behavior of the neutron matter $EOS$, of great relevance 
for the neutron star properties. Actually this substantially modified
parametrization was mainly motivated by a very unpleasant feature
in the spin channel of the previous Skyrme forces, the collapse of
polarized neutron matter, see discussion in \cite{kw94,lyon97}.

A related interesting effect can be seen on the neutron/proton
effective masses.  For an asymmetry $\beta \equiv \frac{N-Z}{A} = 0.2$
(the $^{197}Au$ asymmetry)  
we can get a splitting up to the order of $15\%$ at normal density $\rho_0$,
and increasing with baryon density. 

The sign itself of the splitting is very instructive.
Passing from Skyrme to Skyrme-Lyon we see a dramatic inversion in
the sign of the $n/p$ mass splitting. The Lyon forces predict    
in $n$-rich systems a neutron effective mass always smaller than 
the proton one.
We will come back to this point. Here we just note that the same 
result is obtained
from more microscopic relativistic Dirac-Brueckner calculations
\cite{hole01} and in general from the introduction of scalar isovector
virtual mesons in Relativistic Mean Field ($RMF$) approaches 
\cite{liu02,grelin}. At variance, 
non-relativistic Brueckner-Hartree-Fock
calculations are leading to opposite conclusions \cite{zbl99,zllm02}.
 This will be one of the main
questions to address in the reaction dynamics of exotic nuclear systems.

We can expect important effects
on transport properties ( fast particle emission, collective flows, 
resonance and particle production around the threshold)
 of the dense and asymmetric $NM$ that will
be reached in Radioactive Beam collisions at intermediate energies.

This is the aim of the present paper. We focus our attention on the
isospin dependence of collective flows. We perform collision simulations
based on microscopic kinetic equation models. We start from realistic effective
interactions widely used for symmetric systems. We test very different 
parametrizations in the momentum dependence of the isovector channel,
taking care that the symmetry energy, including its density dependence,
will be not modified. We study in particular the transport effect of
the sign of the $n/p$ effective mass splitting $m^*_n-m^*_p$ in
asymmetric matter at high baryon and isospin density.

We see very interesting $isospin-MD$ effects on $n/p$ transverse
and elliptic flows in semicentral $Au+Au$ collisions at $250~AMeV$
beam energy. The importance of more exclusive data, with a good selection 
of the transverse momenta of the emitted particles, is stressed.

In Sect.2 we present details of the chosen effective interactions.
In Sect.3 the collective flow results are discussed and finally
in Sect.4 some conclusions are drawn with relative perspectives. 

\section{The choice of the effective interactions}

 The
phenomenological non-relativistic effective $EOS$ used in this work was 
first introduced by I.Bombaci et al. \cite{bom95,bom01} in a general form 
suitable for astrophysical
and heavy ion physics applications. We have modified the isovector
part in order to get two new parametrizations with different
$n/p$ effective mass splittings while keeping the same symmetry energy
at saturation, including a very similar overall density dependence.

\begin{figure}[b]
\begin{picture}(0,0)
\put(1.4,3.9){\mbox{\includegraphics[width=2.8cm]{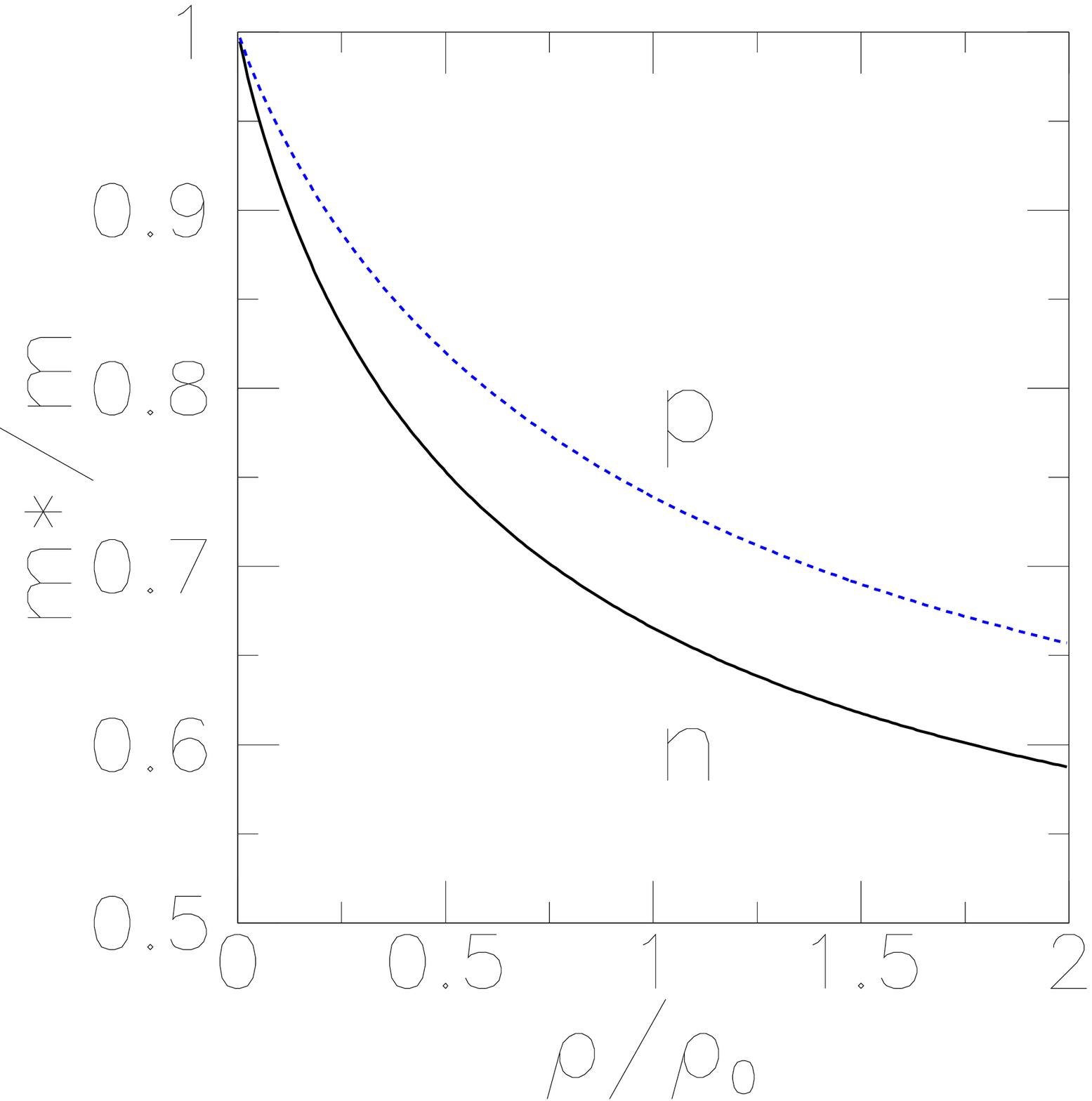}}}
\end{picture}
\includegraphics[width=7.3cm]{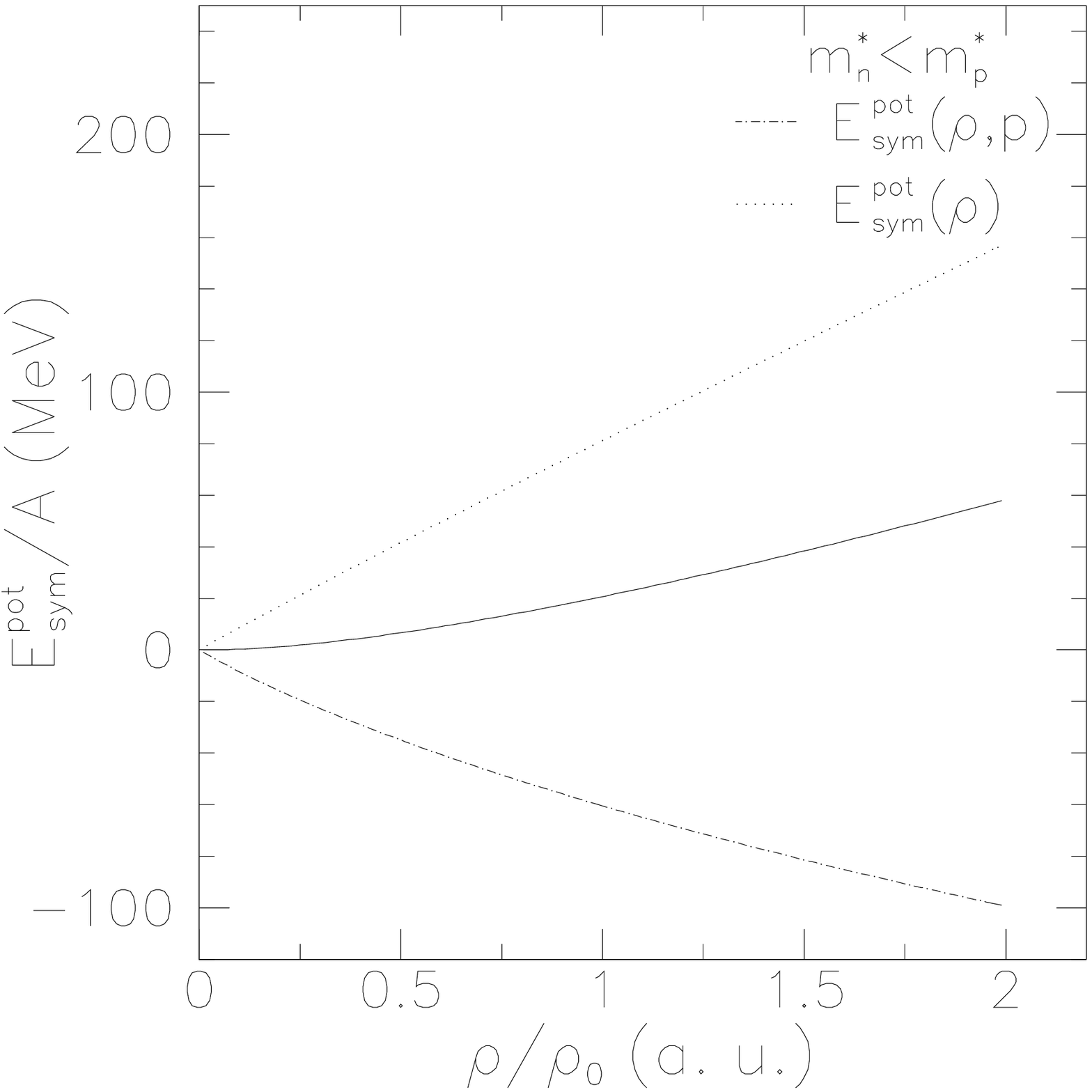}
\includegraphics[width=7.3cm]{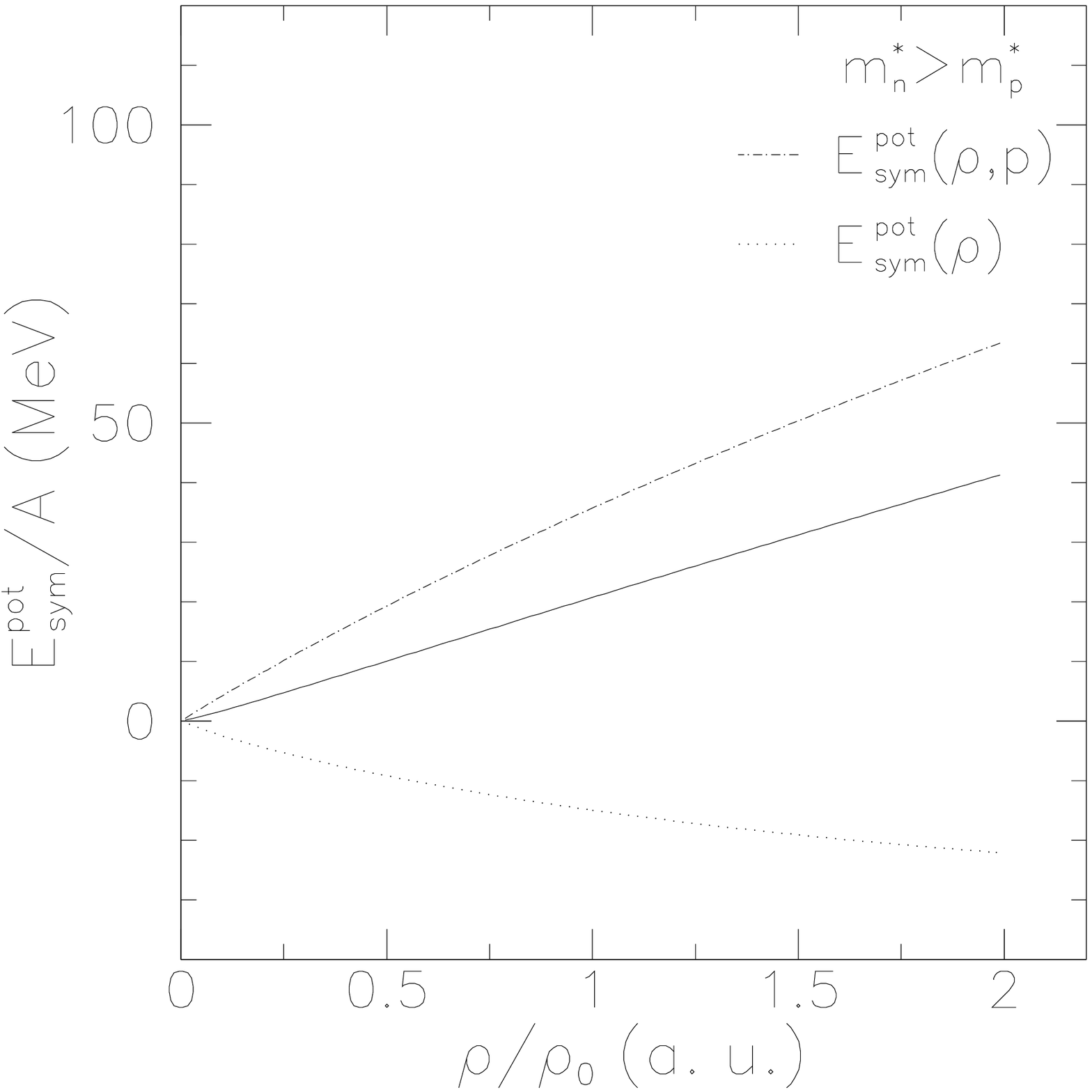}
\begin{picture}(0,0)
\put(8.7,4.3){\mbox{\includegraphics[width=2.8cm]{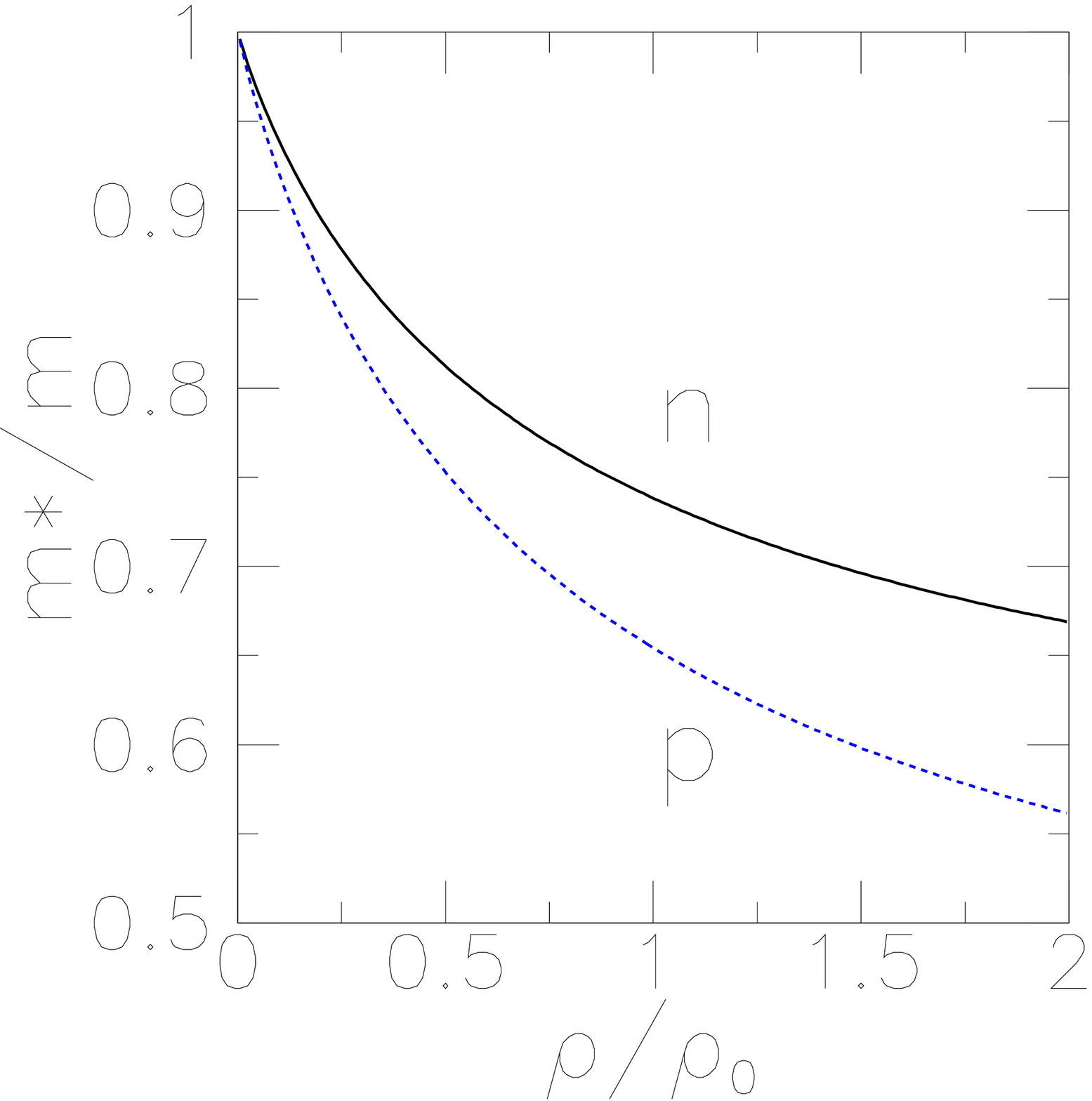}}}
\end{picture}
\caption{Potential symmetry energy as a function of density (solid
line) for $m^*_n<m^*_p$ (left) and $m^*_n>m^*_p$ (right). Dotted
lines refer to density-dependent contributions, dashed-dotted to
momentum dependent ones. Small panels indicate the corresponding
behaviour of proton and neutron effective masses as a function of
density, with $\beta=0.2$.} \label{fig1}
\end{figure}

We start writing down the
energy density as a function of the asymmetry parameter
$\beta \equiv \frac{N-Z}{A}$:
\begin{eqnarray}\label{densene}
\varepsilon(\rho_n,\rho_p)      &=& \varepsilon_{kin}+\varepsilon_A
+\varepsilon_B+\varepsilon_{MD}\\
&&\nonumber\\
\varepsilon_{kin}(\rho_n,\rho_p) &=& \ienne{\frac{\hbar^2}{2m}k^2}
+\ipi{\frac{\hbar^2}{2m}k^2}\nonumber\\
&&\nonumber\\
\varepsilon_A(\rho,\beta)     &=& \frac{A}{2}\frac{\rho^2}{\rz}
-\frac{A}{3}\frac{\rho^2}{\rz}\umd{0}\beta^2\nonumber\\
&&\nonumber\\
\varepsilon_B(\rho,\beta)     &=& \frac{B}{\sigma+1}
\frac{\rho^{\sigma+1}}{\rz^\sigma}-\frac{2}{3}\frac{B}{\sigma+1}
\frac{\rho^{\sigma+1}}{\rz^\sigma}\umd{3}\beta^2\nonumber\\
&&\nonumber\\
\varepsilon_{MD}(\rho_n,\rho_p,\beta)     &=& C\frac{\rho}{\rz}
(\inew{n}+\inew{p})+\frac{C-8z_1}{5}\frac{\rho}{\rz}
\beta(\inew{n}-\inew{p})\nonumber
\end{eqnarray}
where the integrals $\mathcal{I}_\tau(k,\Lambda)=\itau$
include the momentum dependent part of the mean
field $g(k,\Lambda)= \qd{ 1+\td{\frac{k-<k>}{\Lambda}}^2 }^{-1}$;
the subscript $\tau=n,p$ stands for neutrons
 and protons respectively; $\rz=0.16\,fm^{-3}$ is the normal
 density of nuclear matter.
We refer to this parametrization as the $BGBD-EOS$. For symmetric
nuclear matter ($\beta=0$),
the energy density Eq.(\ref{densene}) reduces to the parametrization
proposed by Gale, Bertsch and Das Gupta ($GBD$ interaction,
\cite{GBD}). The parameters $A,\,B,\,C,\,\sigma$ and $ \Lambda$
take the same values as in \cite{GBD} ($A=-144 \,MeV,\, B=203.3
\,MeV,\, C=-75 \,MeV,\,$ $\sigma=\frac{7}{6},\, \Lambda=1.5
\,p_F^{(0)}$, where $p_F^{(0)}$ is the Fermi momentum at normal
density), and provide a soft EOS for symmetric matter, with a
compressibility $K_{NM} \simeq 210~MeV$. 

\begin{figure}[b]
\centering
\includegraphics[width=8cm]{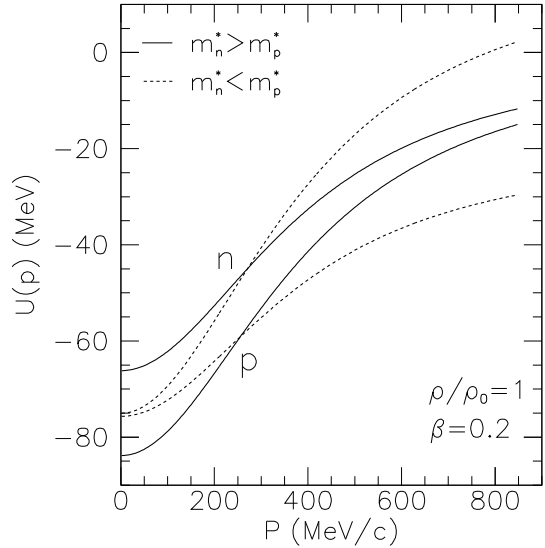}
\caption{Mean field potential as a function of momentum at normal
density, for an asymmetry $\beta=0.2$. Solid lines refer to the
case $m^*_n>m^*_p$, dashed lines to $m^*_n<m^*_p$.} \label{fig2}
\end{figure}

The value of  $z_1$ sets the
strength of the momentum dependence ($MD$) in the isospin channel;
the remaining parameters $x_0\,$ and $x_3$
can be set to fix the symmetry energy.

Taking the functional derivative of the energy density
with respect to the distribution function $f_\tau$,
we obtain, apart from the kinetic term $\varepsilon_{kin}$,
the mean field potential for neutrons and protons:

\begin{eqnarray}
U_\tau(k;\rho,\beta)&=&
A\ra+B\ra^\sigma-\frac{2}{3}(\sigma-1)\frac{B}{\sigma+1}
\umd{3}\ra^{\sigma}\beta^2
\nonumber\\
&&\pm \qd{-\frac{2}{3}A \umd{0}\ra -
\frac{4}{3}\frac{B}{\sigma+1}\umd{3}\ra^{\sigma}\,}\beta\\
&& +\frac{4}{5\rz} \left\{ \frac{1}{2} (3C-4z_1) \inew{\tau}
+ (C+2z_1) \inew{{\tau^{\prime}}}\right\}\nonumber\\
&&+ \td{C \pm \frac{C-8z_1}{5}\beta} \ra g(k)
\nonumber
\end{eqnarray}

where the subscripts in the integrals are $\tau \neq \tau^\prime$;
the upper signs refer to neutrons, the lower ones to protons.
Protons and neutrons in asymmetric matter  experience different
interactions; when $\beta=0$ we return to the $GBD$ mean field. It
is known that a mean field $MD$ implies the idea of an in-medium
reduction of nucleon mass and leads to the introduction of
effective mass \cite{ber88}.

The last $BGBD$ term
includes, besides the usual $GBD$ momentum dependence, an
isospin-dependent part from which we get different effective
masses for protons and neutrons. In fact, the effective mass has
the following definition:

\begin{equation}
\frac{m^*_\tau}{m}=\left\{1+
\frac{m}{\hbar^2k}\frac{dU_\tau}{dk}
\right\}
^{-1}_{k=k_F^{\left[ \tau \right]}}
\label{mstar}
\end{equation}

and we see that mass splitting is
determined not only by different
momentum dependence of mean field,
but also by Fermi momenta of neutrons and protons.
In our case we get:
$$
\frac{m^*_\tau}{m}=\left\{1+\frac{-\frac{2m}{\hbar^2} \frac{1}{\Lambda^2}
\td{
 {C \pm \frac{C-8z_1}{5}\beta}}
 \frac{\rho}{\rho_0}}{
\left[ 1+ \left( \frac{k_{F0}}{\Lambda}
\right)^{^2}   (1 \pm \beta)^{^{(2/3)}}
(\frac{\rho}{\rho_0})^{^{(2/3)}}\right]^2}  \right\}^{-1}
$$
In order to investigate the effects of mass splitting on
non-relativistic heavy ion collisions
we choose two sets of parameters (shown in Table \ref{masses})
which give opposite splitting, but quite similar behaviour of the
symmetry energy.

\begin{table}[htb]
\begin{center}
\begin{tabular}{|c|c|c|c|}
\hline
mass splitting & $z_1$  & $x_0$  & $x_3$\\
\hline
$m^*_n>m^*_p$  & -36.75 & -1.477  & -1.01\\
\hline
$m^*_n<m^*_p$  & 50     & 1.589 & -0.195\\
\hline
\end{tabular}
\caption{Values of the parameters
$z_1,\,x_0$ and $x_3$ for opposite mass splitting, giving the same
$E_{sym}(\rz)=33\, AMeV$.}
\label{masses}
\end{center}
\end{table}

Assuming the
usual parabolic law for the EOS of asymmetric nuclear matter:
$$\frac{E}{A}(\rho,\beta)=\frac{E}{A}(\rho)
+\frac{E_{sym}}{A}(\rho)\beta^2$$ we fix the same symmetry energy at
the saturation density $E_{sym}(\rz)=33\ AMeV$.

Figure \ref{fig1} shows the potential part of the symmetry energy as a
function of density (solid line)
for the two choices of mass splitting. We remark the very similar
overall density dependence.
The dashed-dotted  and dotted lines indicate respectively the
contributions from the momentum dependent
and density-dependent part
of the $EOS$. A change in the relative sign of
mass splitting is related to opposite behaviours of these two
contributions, exactly as already noticed in the Introduction for
the Skyrme-like forces.
Small panels on the top left of each graph illustrate the
corresponding mass splitting as a function
of density for an asymmetry $\beta=0.2$ (the $^{197}Au$ asymmetry).

Let us discuss now the relation between effective mass and
momentum dependence. From the definition Eq.(\ref{mstar}) we see that
the effective mass is inversely proportional to the slope of mean
field at the Fermi momentum. Then, for nucleons with smaller effective
masses the potential will be more repulsive at momenta higher than the
Fermi one, and more attractive at low momenta. Mean field potential at
normal density as a function of momentum  is shown in Figure
\ref{fig2} for a choice $\beta=0.2$. As we see, the parametrizations 
used in this work give rise to opposite behaviors for low and high
momentum particles. This is a very general feature, due to the
meaning itself Eq.(\ref{mstar}) of effective mass. As we will
see in the reaction dynamics this behavior can give rise
to some compensation effects between low and high momenta
contributions. All that will lead to an expected larger 
$isospin-MD$ sensitivity of more exclusive measurements,
in particular with a tranverse momentum selection of the
nucleons emitted for a given rapidity.

\section{Effective Masses and Collective Flows}

\begin{figure}[b]
\centering
\includegraphics[width=8cm]{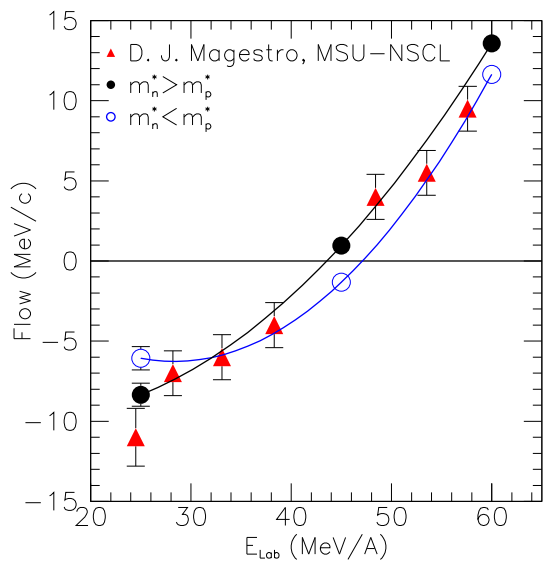}
\caption{Balance energy in $Au+Au$ collisions for intermediate
impact parameters. Solid circles refer to the
case $m^*_n>m^*_p$, empty circles to $m^*_n<m^*_p$.
Triangles represent the data from ref.\cite{mag00}} \label{fig3}
\end{figure}

Collective flows in heavy ion reactions have been widely used to
investigate properties of nuclear matter, such as compressibility
or in-medium cross sections \cite{bao99}, isospin dependence 
\cite{sca99,bao00,bsz01}
and to find out signals of the mean field $MD$ (\cite{zha94}, \cite{dan00}). 
The problem of momentum dependence in the isospin
channel is still open, and recently some possible effects
at relativistic energies have been studied \cite{gre03}. 
Intermediate energies are
important in order to have high momentum particles and to test regions
of high baryon (isoscalar) and isospin (isovector) density during the
reactions dynamics.
Now we
present some qualitative features of the dynamics in heavy ion
collisions related to the splitting of nucleon effective masses.

Our simulations are performed by means of the Boltzmann-Nordheim-Vlasov $BNV$ 
transport code from
Ref. \cite{enzo}, implemented with a $BGBD-like$ mean field, and using
free (isospin dependent) isotropic cross sections in the collision term. 
We focus on
semi-peripheral $Au+Au$ collisions at the energy of $250\, AMeV$ and we
evaluate transverse and elliptic flows of neutrons and protons.
 We consider as emitted particles those resulting in density
regions below $1/8 \,\rz$ during the reaction. We have checked that
the results are not depending on the size of the box where the simulation is 
performed and on the choice of the freeze-out time.


These flow observables can be seen
respectively as the
first and second coefficients from the Fourier expansion of the
azimuthal distribution:
$$\frac{dN}{d\phi}(y,p_t)=1+V_1cos(\phi)+2V_2cos(2\phi)$$
where $p_t=\sqrt{p_x^2+p_y^2}$ is the transverse momentum and $y$
the rapidity along beam direction. 

\subsection{Tranverse flows}

The transverse flow can be
expressed as: 
$$V_1(y,p_t)=\langle \frac{p_x}{p_t} \rangle$$ 
It provides information on the azimuthal anisotropy of the transverse 
nucleon emission
and has been recently used to study the $EOS$ and cross section
sensitivity of the balance energy \cite{bao99}.

Actually the evaluation of the balance energy for the proton more inclusive
{\it Directed Flow}, integrated over all transverse momenta
see \cite{west98,sca99}, in $Au+Au$ collisions represents a very good test
for our effective interactions at lower energies.
The results are shown in Fig.\ref{fig3} and compared with NSCL-MSU
data. Our proton directed flows are calculated for semicentral collisions,
i.e. $b_{red} = 0.5$ \cite{bred}.

Both parametrizations are well reproducing the experimental balance 
energy region. As expected we see a slightly earlier balance of the flow
in the $m^*_n>m^*_p$ case due to some more repulsion for fast protons
but the effect is hardly appreciable at these energies. We will look
more carefully at these effects with increasing beam energy and for more
exclusive observables.

\begin{figure}[t]
\includegraphics[width=7cm]{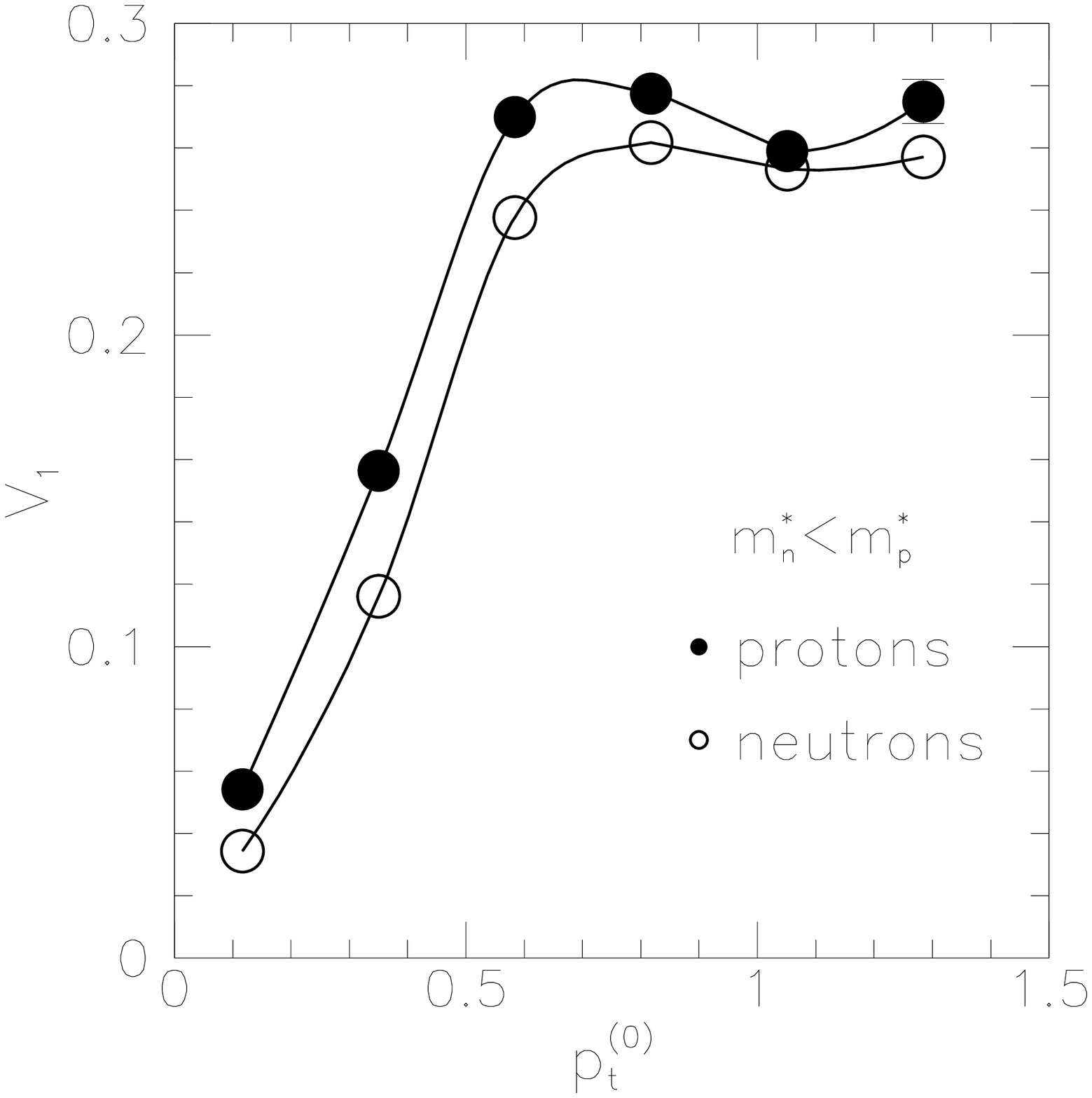}
\includegraphics[width=7cm]{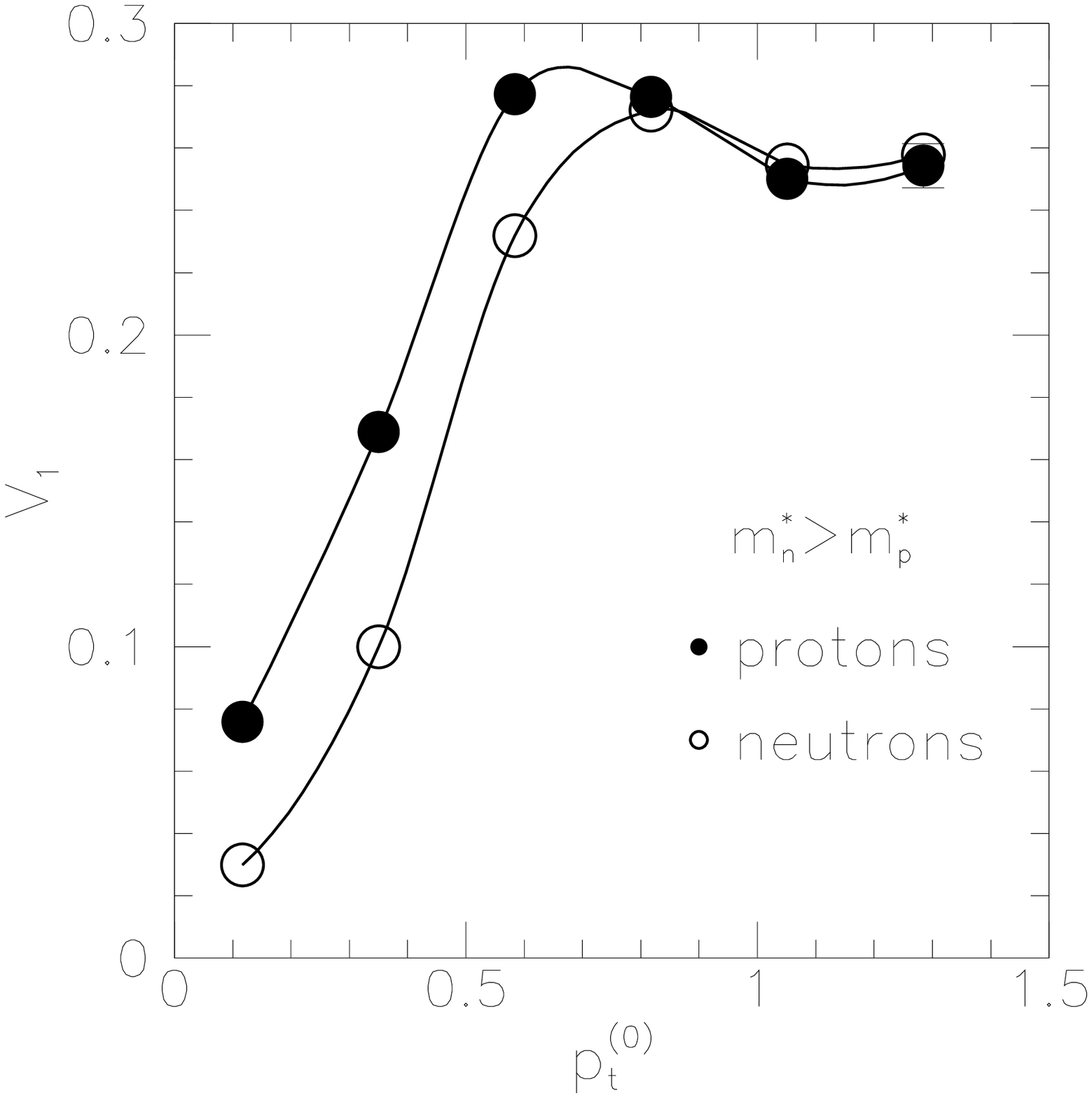}
\caption{Transverse flow of protons and neutrons
for Au+Au reactions at $250\, AMeV$,
$b/b_{max}=0.5$, in the rapidity interval
$0.7\leq |y^{(0)}| \leq 0.9$.}
\label{fig4}
\end{figure}

We will present now $n/p$ transverse flow results for semicentral
$^{197}Au+^{197}Au$ collisions at $250~AMeV$ beam energy, for
particles in regions of relatively high rapidities.
It is useful to work with momenta and rapidities normalized to
projectile ones in the center of mass system ($cm$), defined as
$y^{(0)} \equiv (y/y_{proj})_{cm}$ and
$p^{(0)}_t \equiv (p_t/p_{proj})_{cm}$.
Results of our calculations 
are illustrated in Fig. \ref{fig4}, where the left and
right panels illustrate the two
cases $m^*_n<m^*_p$ and $m^*_n>m^*_p$ respectively.

\begin{figure}[b]
\centering
\includegraphics[width=8cm]{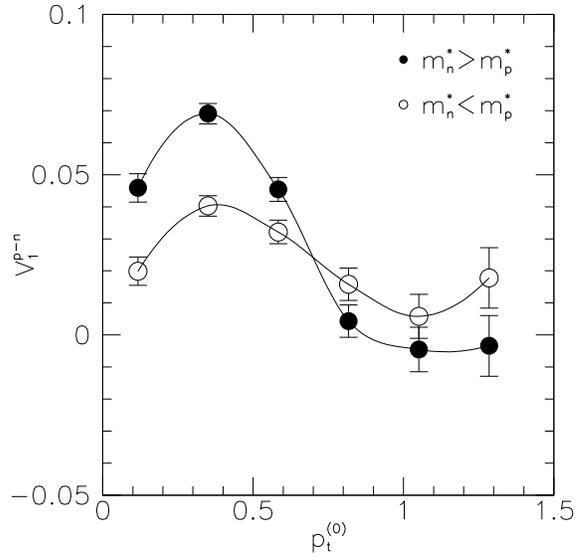}
\caption{$p_t$ dependence of the difference between neutron and proton 
transverse flows in $Au+Au$ 
collisions, same energy, centrality and rapidity selection as in 
Fig.\ref{fig4}. Solid circles refer to the
case $m^*_n>m^*_p$, empty circles to $m^*_n<m^*_p$.} \label{fig5}
\end{figure}

High rapidity selections allow us to find effects of the mean field
$MD$, which becomes more important with increasing momenta. Indeed
the different behaviour of mean field depending on mass splitting
is evident all over the range of transverse momentum shown here.
We emphasize that the average momentum of these particles is
generally beyond the Fermi momentum. Indeed a rough estimate in the center
of mass system
gives already for the beam parallel component
$$p_z\approx m_0 \cdot 0.8(y_{proj})_{cm} \simeq 260-280\,
MeV/c$$ 
In this momentum region a smaller effective mass
determines greater repulsive interaction. When $m^*_n<m^*_p$
neutrons feel greater repulsion than protons and their deflection
in the reaction plane for $p_t^{(0)}\leq 0.6$ is close to
protons, where the Coulomb field is also acting. 
At variance in the case $m^*_n>m^*_p$ the proton flow is greatly
enhanced if compared with neutrons.

At high transverse momentum we must take into account the increase
of the out-of-plane component, so in this case increased repulsion
determines a greater reduction of transverse flow. It causes a
crossing between proton and neutron flow in the case
$m^*_n>m^*_p$. This behaviour is clearly seen if we look at the
different slope of the curves in  Fig. \ref{fig4} (right panel) 
around $p_t^{(0)}\geq
0.6$. 

A quite suitable way to observe such effective mass splitting effect
on the tranverse flows is to look directly at the difference between
neutron and proton flows 
$$V_1^{p-n}(y,p_t) \equiv V_1^{p}(y,p_t) -  V_1^{n}(y,p_t)$$   
shown in Fig.\ref{fig5}.

Now the variation due to the mass splitting choice is quite evident in 
the whole $p_t$ range of emitted 
nucleons. Since the statistics is much smaller at high $p_t$'s (large
error bars in the evaluation) the effect should be mostly observed in the
$p_t^{(0)} \leq 0.5$ range.

\subsection{Elliptic flows}

\begin{figure}[b]
\centering
\includegraphics[width=8cm]{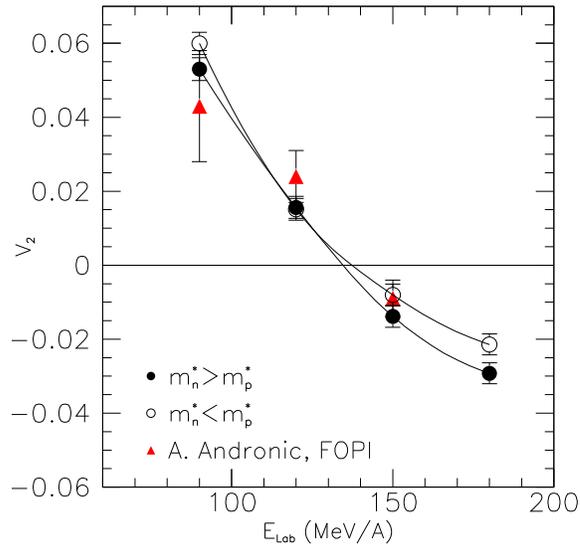}
\caption{Energy dependence of the elliptic flow in $Au+Au$ 
semicentral collisions at mid-rapidity, $|y^{(0)}| \leq 0.1$,
 integrated over the $p_t^{(0)} \geq 0.8$ range. $FOPI$ data with
similar selections, see ref. \cite{fopi99},
are given by the full triangles. 
Solid circles refer to the
case $m^*_n>m^*_p$, empty circles to $m^*_n<m^*_p$.} \label{fig6}
\end{figure}

The competition between in-plane and out-of-plane emissions can be
pointed out by looking at the elliptic flow, which has the
following expression:
$$V_2(y,p_t)=\langle \frac{p_x^2-p_y^2}{p_t^2} \rangle$$
The sign of $V_2$ indicates the azimuthal anisotropy of emission:
particles can be preferentially emitted either in the reaction
plane ($V_2>0$) or out-of-plane (squeeze-out, $V_2<0$). 

Also in this case a good check of our effective interaction choices
is provided by some more inclusive data in the medium energy range.
In Fig.\ref{fig6} we report a comparison with the $FOPI-GSI$ data in the
interesting beam energy region of the change of sign of the elliptic flow for
protons, integrated over large transverse momentum contributions.
The agreement is reasonable for both parametrizations. Also in this
case we see a slightly larger overall repulsion (earlier zero-crossing) 
for the $m^*_n>m^*_p$ case.

\begin{figure}[t]
\includegraphics[width=7cm]{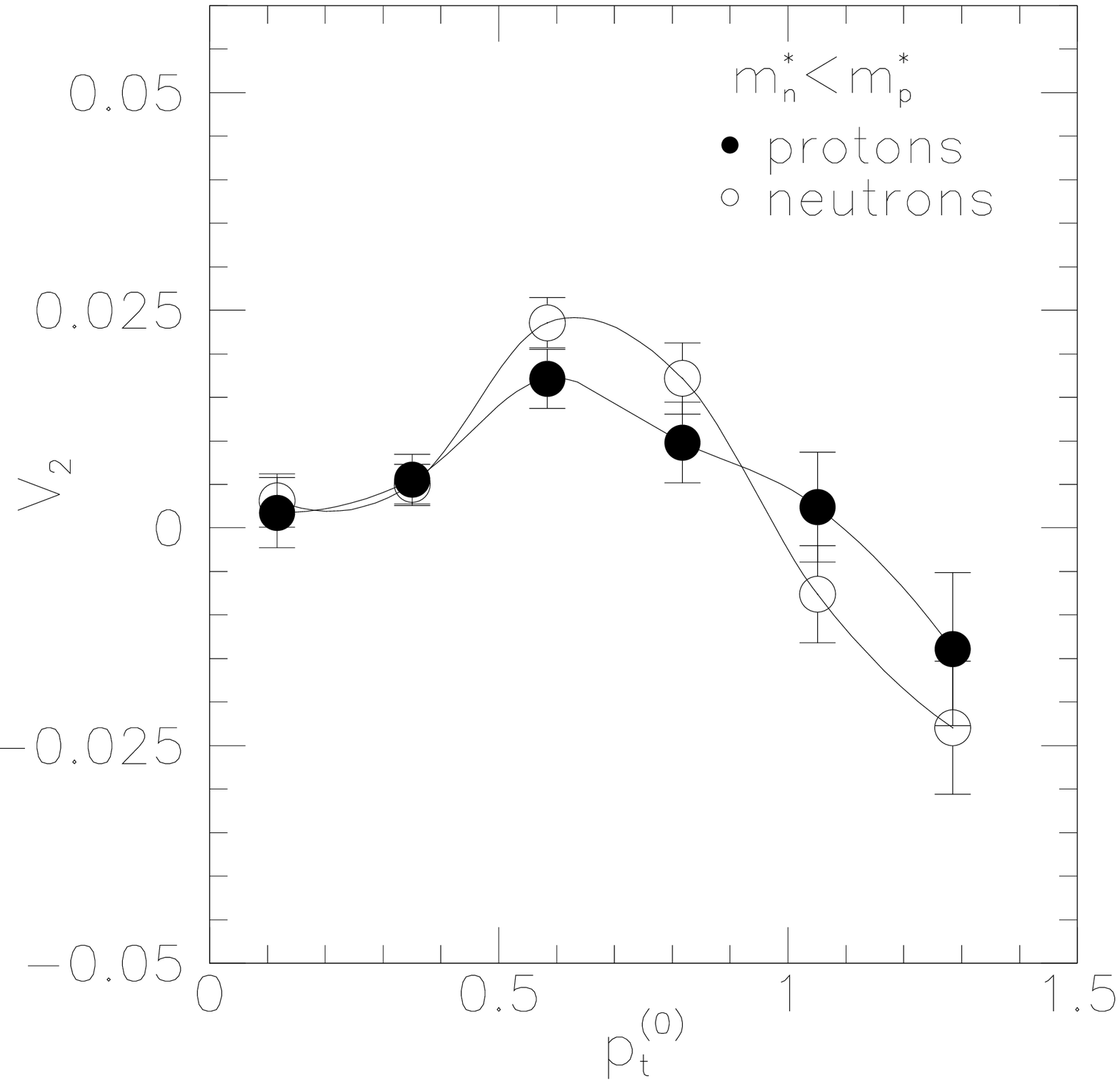}
\includegraphics[width=7cm]{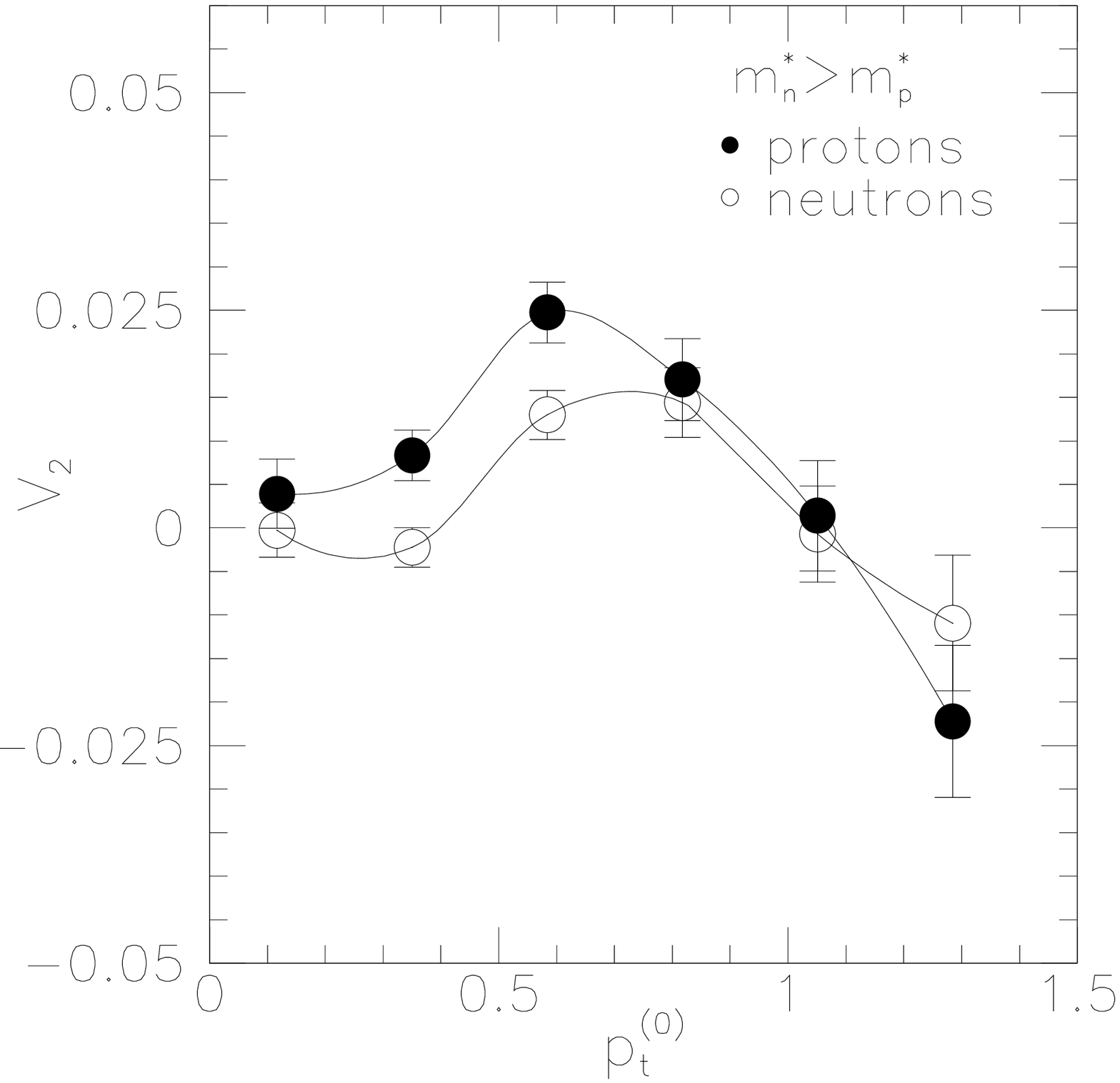}
\put(-12.8,2.85){\mbox{\line(1,0){5}}}
\put(-5.75,2.85){\mbox{\line(1,0){5}}} \caption{Elliptic flow of
protons and neutrons in Au+Au reaction at $250\, AMeV$,
$b/b_{max}=0.5$, in the rapidity interval $0.7\leq |y^{(0)}| \leq
0.9$.} \label{fig7}
\end{figure}

Our results on the $p_t$ dependence of the neutron/proton
elliptic flows in $Au+Au$ reactions at $250~AMeV$ are shown in Fig.
\ref{fig7}. We can see that the $p_y$ component rapidly grows at
momenta $p_t^{(0)}>0.6$, causing the flows to fall down towards
negative values. At low momenta, emitted particles undergo a
deflection in the reaction plane, more pronounced for nucleons
with smaller effective mass.
 In fact, comparing the two panels of Fig.\ref{fig7} we see that
 around the maximum in-plane deflection 
$0.4\lesssim p_t^{(0)}\lesssim 0.8$ neutron and proton flows are inverted. We
also find different slopes for protons and neutrons around $p_t^{(0)}\geq
0.7$, depending on
the relative sign of mass splitting, in agreement with the
previous results in the transverse flows.
 Around $p_t^{(0)} \simeq 1$,
i.e. projectile momentum in the $cm$,  the elliptic flow finally takes 
negative
values, with slight differences for the two parametrizations.

\begin{figure}[b]
\centering
\includegraphics[width=8cm]{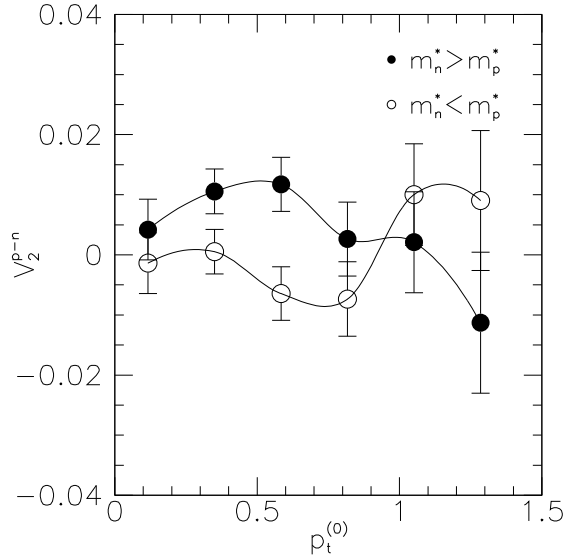}
\caption{$p_t$ dependence of the difference between neutron and proton 
transverse flows in $Au+Au$ 
collisions, same energy, centrality and rapidity selection as in 
Fig.\ref{fig7}. Solid circles refer to the
case $m^*_n>m^*_p$, empty circles to $m^*_n<m^*_p$.} \label{fig8}
\end{figure}

Also in this case the best observable to look at is the $p_t$
dependence of the difference between neutron and proton elliptic flows
$$V_2^{p-n}(y,p_t) \equiv V_2^{p}(y,p_t) -  V_2^{n}(y,p_t)$$ 
shown in the Fig.\ref{fig8}.

Again the difference is quite evident in the whole $p_t$ range of emitted 
nucleons. We note again that the statistics is much smaller at high $p_t$'s 
(large error bars in the evaluation). Anyway the mass splitting effect 
appears very clearly  in the low
$p_t^{(0)} \leq 0.5$ range where the results are more reliable.

As a general comment we stress the importance of more exclusive data,
with a good $p_t$ selection. Indeed from Figs. \ref{fig4}, \ref{fig5}
(for $V_1$) and Figs. \ref{fig7}, \ref{fig8} (for $V_2$) we clearly see
that more inclusive $V_1(y)$, $V_2(y)$ data, integrated over
all $p_t$'s of the emitted nucleons at a given rapidity,
will appear to be less sensitive to $isospin-MD$ effects. As already discussed
from general features of the mean field $MD$ we expect a kind of compensation
between low and high $p_t$ contributions.

\begin{figure}[b]
\includegraphics[width=7cm]{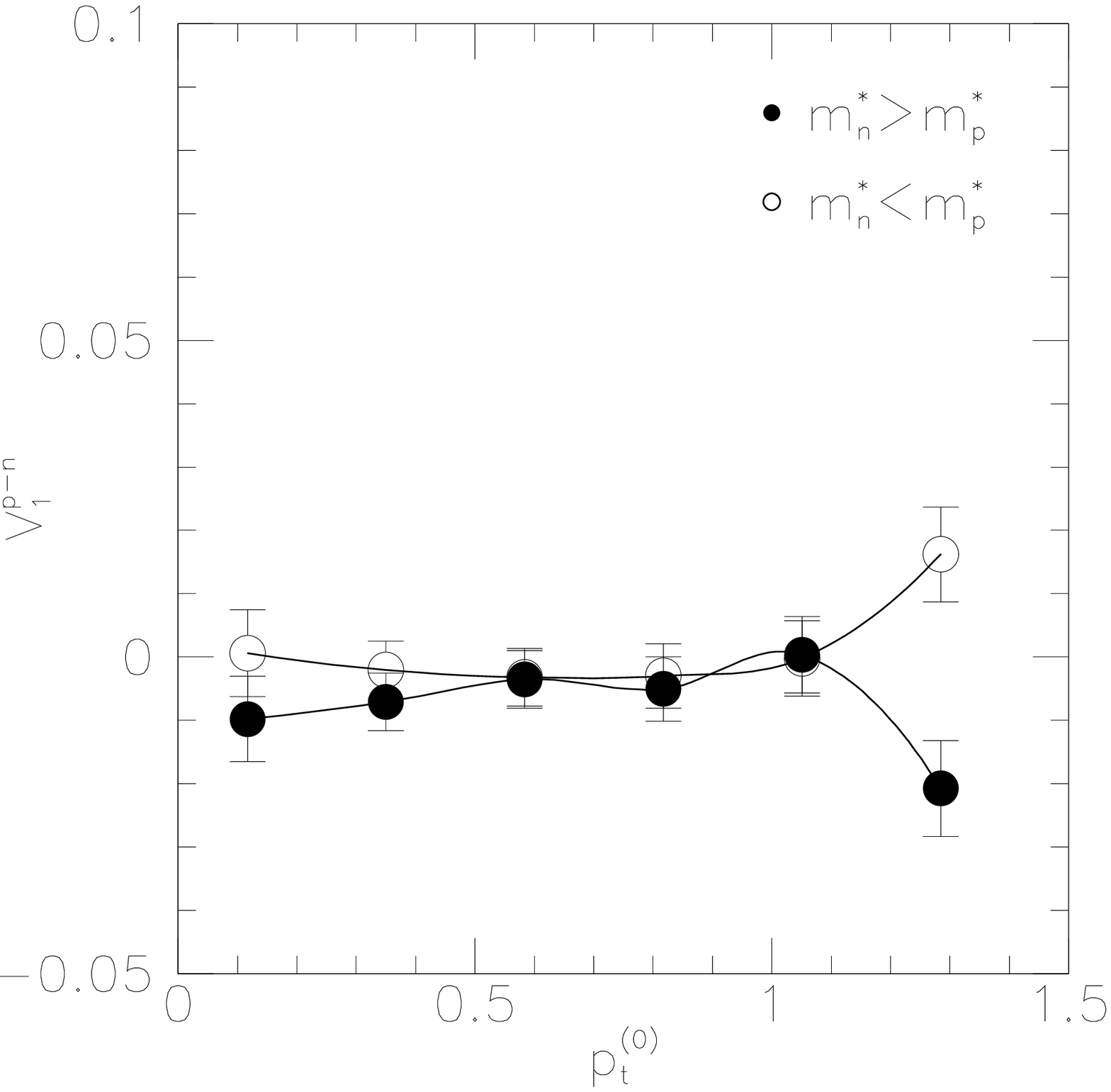}
\includegraphics[width=7cm]{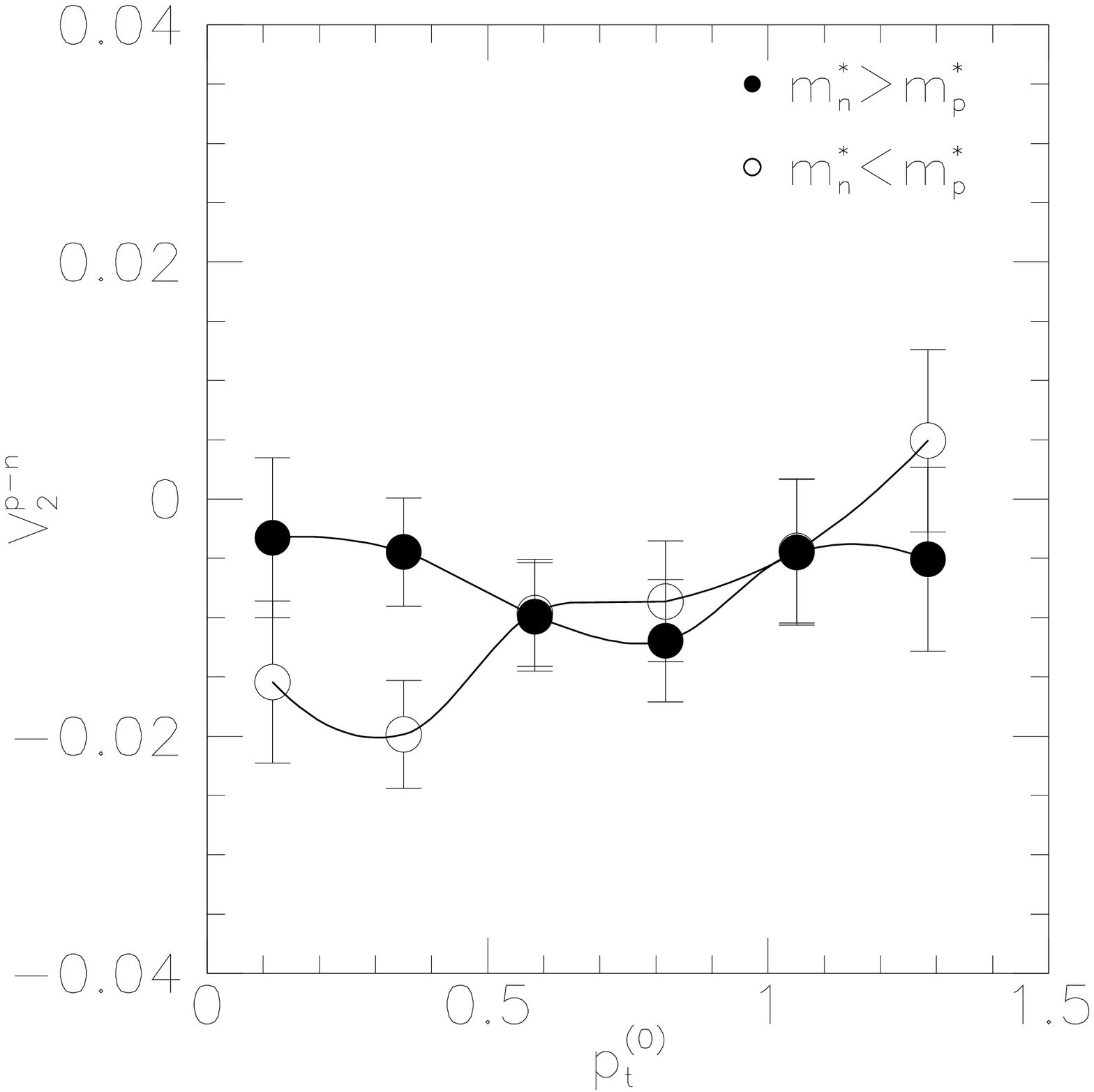}
\caption{$p_t$ dependence of the difference between neutron and proton
transverse (left panel) and elliptic flow (right panel)
in $Au+Au$ collisions at $250\, AMeV$,
$b/b_{max}=0.5$, in the rapidity interval
$0\leq |y^{(0)}| \leq 0.25$. Solid circles refer to the
case $m^*_n>m^*_p$, empty circles to $m^*_n<m^*_p$.}
\label{fig9}
\end{figure}

\subsection{Low rapidity results}

In order to confirm the interpretation of the previous flow results
we have performed the same analysis for a low rapidity selection
$0\leq |y^{(0)}| \leq 0.25$. Since we are now looking at particles with
overall smaller momenta, the effects of the isospin contributions to the
mean-field-$MD$ should be reduced.

We illustrate this point in
Fig.\ref{fig9}, where the difference between proton and neutron flows 
are plotted  as a function of transverse momentum in the low rapidity region.
We have behaviors quite similar to the corresponding 
Figs.\ref{fig5}, \ref{fig8},
but the effective mass splitting cannot be clearly observed.
 
We can see some indications of a larger attraction felt by low momenta protons
in the case $m^*_n>m^*_p$, see the $p$-solid line of Fig.\ref{fig2}.
This can be observed in the reduction of the proton transverse flow
at low $p^{(0)}_t$, solid circles below the empty ones in the range
$p^{(0)}_t \leq 0.5$. The effect looks interesting, although not
much appreciable because of large uncertainty.

\subsection{Changing the stiffness of the symmetry term}

It is well known that the collective flows for asymmetric systems are 
sensitive to the density dependence of the symmetry term of
the nuclear EOS, in particular at high transverse momentum 
\cite{sca99,bao00,bsz01}.
 
For this reason it is important to be sure that signals previously
discussed are mainly due to the $n/p$ effective mass splitting and 
not strictly 
depending on the choice of the stiffness of the symmetry energy.

\begin{figure}[hb]
\includegraphics[width=7cm]{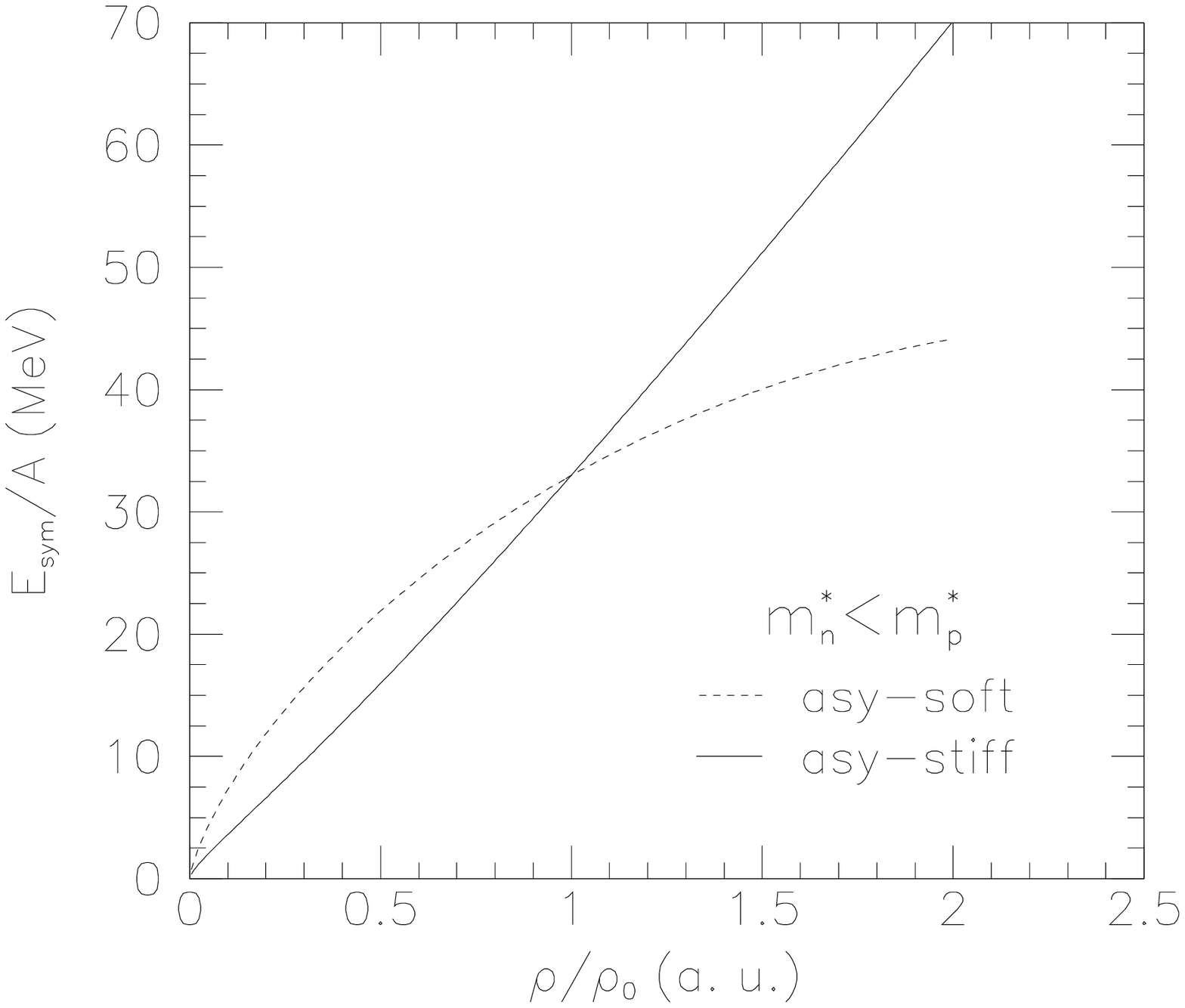}
\includegraphics[width=7cm]{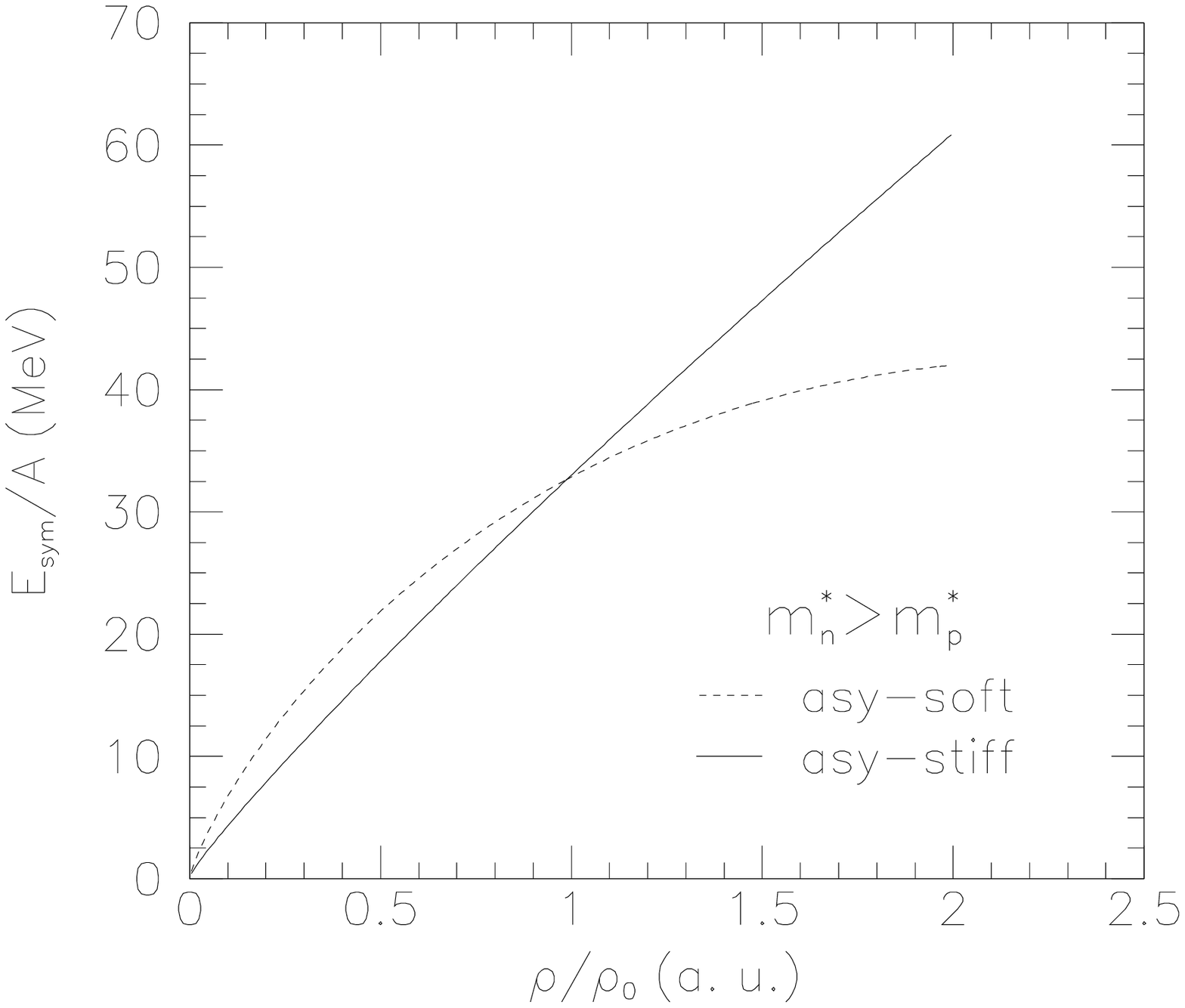}
\caption{Total symmetry energy as a function of density for the two mass
splitting choices used in this work. 
Solid and dashed lines represent respectively 
the $asy-stiff$ and $asy-soft$ equations of state.}
\label{fig10}
\end{figure}

In order to check this point
we have performed some calculations using a symmetry energy with a 
density behaviour very different from the one used before, 
{\it but keeping the same $n/p$-effective mass splittings}. 
We refer to it as 
{\it asy-soft}, see Fig.\ref{fig10}, compared with the {\it asy-stiff} 
previously used (shown also in Fig.\ref{fig1} for the potential part). 
Since aim is to leave the momentum dependence of the symmetry
potential unchanged 
and to modify only the density dependence, 
we take the same $z_1$ values used before and
choose new $x_0$ and $x_3$ values. In this way we can generate a softer 
symmetry energy at high density, of course
with the same saturation value $E_{sym}(\rz)=33\,AMeV$. Table \ref{asy-soft}
reports the two sets of new parameters. 

\begin{table}[tb]
\begin{center}
\begin{tabular}{|c|c|c|c|}
\hline
mass splitting & $z_1$  & $x_0$  & $x_3$\\
\hline
$m^*_n>m^*_p$  & -36.75 & 0.11  & 0.21\\
\hline
$m^*_n<m^*_p$  & 50     & 4.449 & 2.\\
\hline
\end{tabular}
\caption{Values of the parameters
$z_1,\,x_0$ and $x_3$ for opposite mass splitting, giving a similar 
soft simmetry term.}
\label{asy-soft}
\end{center}
\end{table}

In Fig. \ref{fig10} we show the
total (potential plus kinetic) symmetry energy in the two cases,
$m^*_n<m^*_p$ (left panel) and $m^*_n>m^*_p$ (right panel). Solid and dashed
lines represent respectively the $asy-stiff$ and $asy-soft$ curves. 

\begin{figure}[b]
\includegraphics[width=7cm]{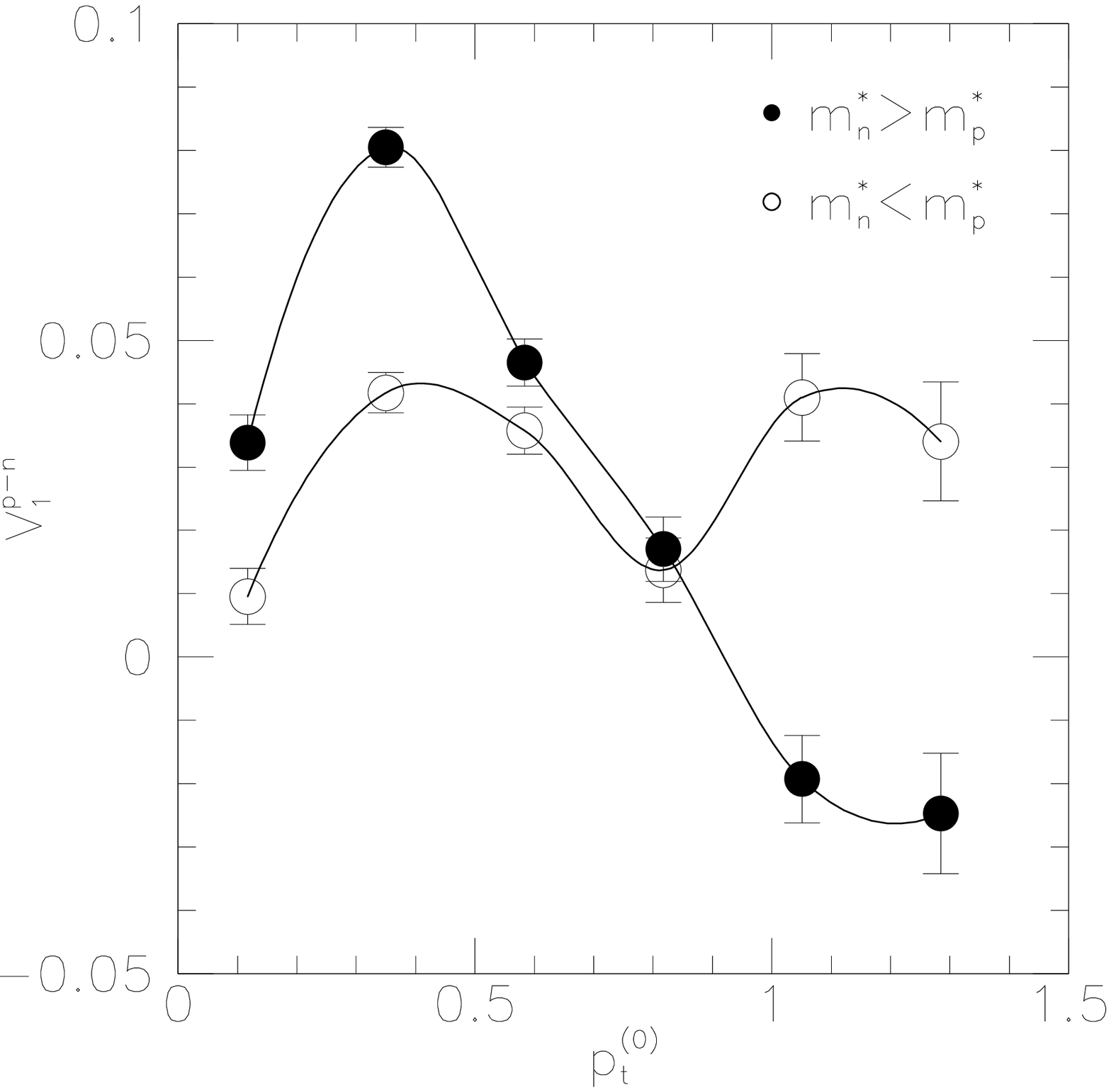}
\includegraphics[width=7cm]{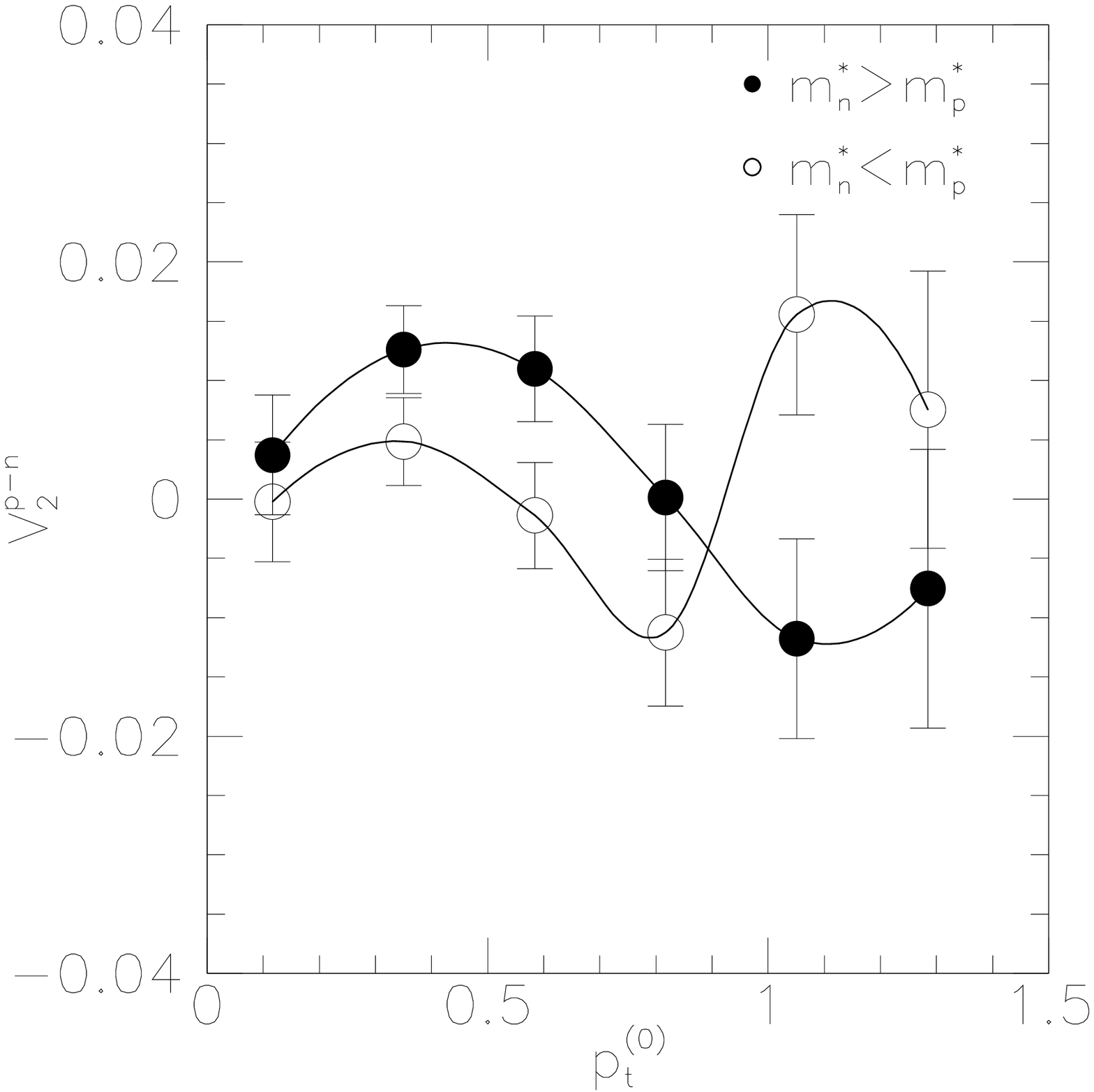}
\caption{$p_t$ dependence of the difference between neutron and proton
transverse (left panel) and elliptic flow (right panel)
in $Au+Au$ collisions at $250\, AMeV$,
$b/b_{max}=0.5$, in the rapidity interval
$0.7\leq |y^{(0)}| \leq 0.9$. Calculations are performed using the $asy-soft$
symmetry energy. Solid circles refer to the
case $m^*_n>m^*_p$, empty circles to $m^*_n<m^*_p$.}
\label{fig11}
\end{figure}

Our results show that while the nucleon flows are depending on
the stiffness of the symmetry term, like in other calculations
\cite{sca99,bao00,bsz01}, the transverse momentum behavior
of the differences keeps the same sensitivity to the effective mass
splittings. 

Fig. \ref{fig11} illustrates the $n/p$-flow differences as a function of
transverse momentum for the same energy, impact parameter and high rapidity
selections used before. Both transverse $V_1^{p-n}(y,p_t)$ and 
elliptic $V_2^{p-n}(y,p_t)$ flow 
differences show very small
variations from the $asy-stiff$ case (i.e. Figs. 
\ref{fig5} and \ref{fig8}) over the whole $p_t^{(0)}$ range, in
particular
in the region  
$p_t^{(0)}\leq0.5$, of interest for the better statistics. 

This result represents an important indication that we are 
selecting experimental observables that
directly probe the isospin effects on the momentum dependent part
of the nuclear $EOS$.

\section{Conclusion and outlook}

We have presented some expected influence of the momentum dependence
in the isovector channel of the effective $NN$ interaction on
the heavy ion reaction dynamics at intermediate energies. 

For a given parametrization, dynamical effects of momentum
dependence in the isospin channel are opposite for neutrons and
protons, so that they can be magnified by looking at the
differences between neutron and proton collective flows. This kind of
measurements requires the detection of neutron flows, but it is very useful
to draw out some information about the relative sign of the effective mass
splitting in a asymmetric nuclear medium.

In spite of the low asymmetry we have found interesting effects
from the study of $Au+Au$ semicentral collisions at $250~AMeV$.
These results can provide some hints for future measurements,
also with unstable beams.

We suggest to look at the flows of particles of relatively large rapidity,
in particular performing more exclusive experiments for various
transverse momentum selections. A compensation is indeed expected
for observables where all $p_t$ contributions are summed up. This
is physically understandable on general grounds due to the different
interplay between low and high momentum mean field repulsion, for the
various effective mass choices. 

These more exclusive flow observables, in particular the 
transverse $V_1^{p-n}(y,p_t)$ and 
elliptic $V_2^{p-n}(y,p_t)$ flow 
differences,
appear to be very selective probes
of the momentum dependence of the isovector part of the nuclear $EOS$.
The observed effects are not changing when we largely modify the
density dependence of the symmetry term while keeping unchanged
the momentum dependence of the symmetry mean field.

Due to the difficulty of the neutron flow detection, good data with similar 
features could be obtained from the flows of light isobars, like
$^3H, ^3He$ and so on. Exclusive data, with a $p_t$ selection, appear
in any case of large importance.

Collective flow measurements seem to represent a very nice exploration of the
highly controversial sign of the $n/p$ effective mass splitting in
asymmetric matter, of large fundamental interest.
Moreover related important effects can be expected in a wide spectrum
of the nuclear physics phenomenology, from the structure of drip-line 
nuclei to the threshold production of resonances and particles in $HIC$ with
radioactive beams.




\begin{thebibliography}{99}
\bibitem{bred} The reduced impact parameter is defined as 
$b_{red} \equiv b/b_{max}$ where $b_{max}$ is the sum of the 
 two nuclear radii.
\end{thebibliography}
\end{document}